\documentclass[aps,pra,superscriptaddress,twocolumn,longbibliography]{revtex4-1}
\usepackage{units}
\usepackage{amsmath}
\usepackage{amsthm}
\usepackage{amssymb}
\usepackage{graphicx}
\usepackage{color}
\usepackage{bbold}
\usepackage[colorlinks=true, urlcolor=blue, citecolor=blue,linkcolor=blue,citebordercolor={1 0 0},linkbordercolor={0 0 1}]{hyperref}

\theoremstyle{plain}

\theoremstyle{plain}

\ifx\proof\undefined
\newenvironment{proof}[1][\protect\proofname]{\par
	\normalfont\topsep6\p@\@plus6\p@\relax
	\trivlist
	\itemindent\parindent
	\item[\hskip\labelsep\scshape #1]\ignorespaces
}{%
	\endtrivlist\@endpefalse
}
\providecommand{\proofname}{Proof}
\fi
\theoremstyle{plain}

\theoremstyle{remark}

\makeatother

\providecommand{\factname}{Fact}
\providecommand{\theoremname}{Theorem}
\providecommand{\claimname}{Claim}
\providecommand{\lemmaname}{Lemma}
\providecommand{\definitionname}{Definition}

\theoremstyle{definition}

\graphicspath{{./Images/},{./ImagesAppendix/}}

\begin{document}
\title{Binary classifiers for  noisy datasets: a comparative study of existing quantum machine learning frameworks and some new approaches}

\author{N. Schetakis}
\thanks{Equal contribution}
\affiliation{School of Electronic and Computer Engineering, Technical University of Crete, Chania, Greece 73100}
\affiliation{ALMA Sistemi Srl, Guidonia (Rome), 00012, Italy}
\email{nsx@alma-sistemi.com}
\author{D. Aghamalyan}
\thanks{Equal contribution}
\affiliation{School of Information Systems, Singapore Management University, 81 Victoria Street, Singapore 188065}
\affiliation{Centre for Quantum Technologies, National University of Singapore, Singapore 117543}
\author{M. Boguslavsky}
\thanks{Equal contribution}
\affiliation{Tradeteq Ltd, London, UK}
\author{P. Griffin}
\thanks{Equal contribution}
\affiliation{School of Information Systems, Singapore Management University, 81 Victoria Street, Singapore 188065}

\begin{abstract}
One of the most promising areas of research to obtain practical advantage is Quantum Machine Learning which was born as a result of cross-fertilisation of ideas between Quantum Computing and Classical Machine Learning. In this paper, we apply Quantum Machine Learning (QML) frameworks to improve binary classification models for noisy datasets which are prevalent in financial datasets. The metric we use for assessing the performance of our quantum classifiers is the area under the receiver operating characteristic curve (ROC/AUC). By combining such approaches as hybrid-neural networks, parametric circuits, and data re-uploading we create QML inspired architectures and utilise them for the classification of non-convex 2 and 3-dimensional figures. An extensive benchmarking of our new FULL HYBRID classifiers against existing quantum and classical classifier models, reveals that our novel models exhibit better learning characteristics to asymmetrical Gaussian noise in the dataset compared to known quantum classifiers and performs equally well for existing classical classifiers, with a slight improvement over classical results in the region of the high noise. 
\end{abstract}
\maketitle

\noindent {\em Introduction.---} %NISQ and types of quantum advantage
Noisy Intermediate-Scale Quantum (NISQ)~\cite{preskill2018quantum,bharti2021noisy,deutsch2020harnessing} devices hold a promise to deliver a practical quantum advantage by harnessing the complexity of quantum systems. Despite being several years away from having fault-tolerant quantum computing\cite{preskill1998fault,gottesman1998theory, shor1996fault}, researchers have been hopeful to achieve this task. Perhaps one of the most exciting breakthroughs in this direction was a demonstration of "quantum supremacy" by Google researchers~\cite{arute2019quantum}, using their programmable superconducting Sycamore chip with 53 qubits, in which single-qubit gate fidelities of 99.85\% and two-qubit gate fidelities of 99.64\% were obtained on average. Here the task of sampling
the output of a pseudo-random quantum circuit was successfully achieved. Quantum Supremacy would imply that a universal quantum computer has the ability to perform certain tasks exponentially faster than a classical computer\cite{harrow2017quantum}. However, it has been argued later that Google's achievement amounted to a demonstration of a quantum advantage but not a practical advantage, in other words, the performed task was not useful for any real-life applications. Another quantum advantage breakthrough experiment has been implemented \cite{zhong2020quantum} utilising a Jiuzhang photonic quantum computer and performing Gaussian boson sampling (GBS) with 50 indistinguishable single-mode squeezed states. Here, quantum advantage has been elucidated in the sampling time complexity of a Torontonian matrix, which has exponential scaling with output photon clicks. However, this experiment demontsrates quantum advantage but fails to demonstrate quantum supremacy as this photonic quantum computer is not programmable.
%QML

One of the most promising areas of research to obtain practical advantage is  Quantum Machine Learning\cite{biamonte2017quantum,wittek2014quantum,schuld2018supervised} which was born as a result of cross-fertilisation of ideas between Quantum Computing~\cite{nielsen2002quantum,preskill1998lecture} and Classical Machine Learning~\cite{goodfellow2016machine,jordan2015machine}. QML in its spirit is similar to classical machine learning but with the main difference being that instead of classical neurons in the layers of a deep neural network, now we have qubits and quantum gates acting on qubits combined with quantum measurements playing the role of the activation function. The elegant field of QML has been providing a new platform for devising algorithms that exhibit quantum speedups. For instance, it has been demonstrated that such basic linear algebra subroutines as solving certain types of linear equations (the quantum version is known in the community as HHL), finding eigenvectors and eigenvalues, principal component analysis (PCA) exhibit exponential speedups compared to their classical counterparts\cite{harrow2009quantum,huang2019near,rebentrost2018quantum,lloyd2014quantum,wiebe2012quantum}. Since we are dealing with a quantum system, one can utilise such quantum resources as coherence, entanglement, negativity, contextuality to leverage towards achieving practical advantage. However, it is still not completely understood what the role of different types of resources is in harnessing practical advantage from available 50-100 qubit noisy devices~\cite{deutsch2020harnessing}.
%3 hurdles of QML -focus on data encoding Lattore, Maria Schuld Book
The three main building blocks of any QML algorithm are data encoding, unitary evolution of the system followed by the state readout performed through the measurement~\cite{schuld2018supervised}. Uploading classical data in the quantum computer is not a trivial task and can account for most of the complexity of the algorithm, determining what kind of speed-ups are feasible. This procedure is called quantum embedding which can be achieved, for instance, with help of "quantum feature  maps"\cite{schuld2019quantum,perez2020data,perez2021one,schuld2020embedding,mitarai2018quantum} which take classical data and map it to the high-dimensional Hilbert space, where one hopes to achieve higher separation between the data classes compared to the original coordinate system. Moreover, one can train the quantum embedding to achieve maximal separation between the data clusters in the Hilbert space (this approach has been coined as "quantum metric learning")\cite{schuld2020embedding,mitarai2018quantum}, paving the way towards constructing faithful quantum classifiers. 

Binary classification is a  ubiquitous task in machine learning. Perhaps the most prominent example is the cat recognition algorithm, which gives a flavour of the power brought by utilising such basic tools as logistic regression combined with deep neural network architectures~\cite{goodfellow2016machine}. Quantum classifiers hold a promise to bring feasible speedups compared to their classical counterparts. Several theoretical proposals combined with actual experimental runs on commercially available backends have been put forward for realising faithful quantum classifiers\cite{schuld2020circuit,schuld2019quantum,farhi2018classification,perez2020data,tacchino2019artificial,cappelletti2020polyadic,wiebe2016quantum,liao2018quantum,schuld2017implementing,tiwari2019towards,blank2020quantum,park2020theory}.
For instance, approaches in Refs.~\cite{blank2020quantum,park2020theory} are inspired by kernel methods used in classical machine learning. Refs.~\cite{schuld2020circuit,schuld2019quantum,farhi2018classification} are combining certain types of quantum embeddings to achieve quantum hybrid neural networks, which are promising candidates for building a faithful classifier. Ref.~\cite{acchino2019artificial} suggests using hypergraph-states\cite{rossi2013quantum}, where the assumption is that such states can lower the circuit depth of the classifier. Refs.\cite{wiebe2016quantum,liao2018quantum} are based on quantum Grover's search algorithm.

%Survey all the proposals review and try to put them down then start introducing your own approach
%Introduce building blacks of your approach and survey those blocks-hybrid neural networks, variational circuit, angle embedding,data-reuploading
%cut to the chase by saying what you actually do we maximize ROC/AUC for 3 different types of hypersurfaces and doing hyperparameter tuning, explain ROC/AUC in one sentence, why it is important? emphasize that you focus on the particular type of datasets that are prevalent in financial datasets, what is your deliverable? software, we study the role of entanglement

In this manuscript, we take a rather pragmatic approach and try to benefit from a plethora of available QML software packages\cite{bergholm2018pennylane,killoran2019strawberry,broughton2020tensorflow,efthymiou2020qibo,kottmann2021tequila}, which grant access to run the quantum circuit in the quantum simulator or an actual hardware (such as IBM Quantum Experience, Amazon Braket, Rigetti Computing, Strawberry Fields). By utilising these tools we provide new software that is particularly well suited for targeting classification problems in the unbalanced and noisy datasets which are prevalent in the financial industry\cite{orus2019quantum}. 

In this paper at first we briefly outline and review three different necessary building block QML architectures for our software package  : hybrid-neural networks~\cite{schuld2020circuit,schuld2019quantum,farhi2018classification}, parametric quantum circuits~\cite{benedetti2019parameterized,cerezo2020variational,bharti2021noisy,funcke2021dimensional} and data-reuploading~\cite{perez2020data,perez2021one}. 

The metric we use for assessing the performance of our quantum classifiers is the area under the receiver operating characteristic curve (ROC/AUC). ROC is a probability curve and AUC represents the degree of separability. In general a good model has AUC close to 1.  We test our FULL HYBRID models and benchmark them against existing QML classifiers and also to the best known classical machine learning counterparts by running simulations on quantum simulators for three different 2-dimensional non-convex surfaces. It is believed that non convex boundaries represent more difficult classification problems as linear regression is bound to fail in this tasks. Then by introducing asymmetrical Gaussian noise we study the resilience of our different approaches to the noise. This kind of study sheds light on learning properties for the amount of noise in the dataset. We also perform systematic hyperparameter tuning by studying how ROC/AUC changes with the number of repeating units in the data-re-uploading approach, number of qubits,  batch size, number of epochs and number of strongly entangling units. We remark, that our binary classifiers can be extended to multi-class classification problems using a one-versus-all approach.
%Our study provides an insight into the role of entanglement in boosting learning capabilities in QML frameworks. We remark, that our binary classifiers can be extended to multi-class classification problems using a one-versus-all approach.
%%%%%%%%%%%%%%%%%%%%%%%%%%%%%%%%%OUTLINE%%%%%%%%%%%%%%%%%%%%%%%%%%%%%%%%%%%%%%%%%%%%%%%%%%%%%%%%%%

The manuscript is organised as follows. In section \ref{setting} we explain what kind of classification problems for 2 and 3-dimensional non-convex surfaces we tackle in the current study. In  Section \ref{Review of existing QML frameworks}, we briefly review the three  building blocks in the QML frameworks which we utilise in the next section \ref{FULLHYBRID} for our novel classification circuit which combines the best features of the building blocks.  In  Section \ref{Comparative study of different frameworks} we benchmark several known QML approaches(including our  QML classifier) along with the best known classical counterparts for binary classification problems. Here we focus, in particular, on prediction grids and ROC/AUC characteristics for assessing the performance of the classifier. Section \ref{conclusions} is devoted to the conclusions and future directions.Finally, in Appendix~\ref{Appendix} material we show some results for 3-dimensional non-convex boundary classification problems and demonstrate the performance of our FLL HYBRID classifiers.
 
%%%%%%%%%%%%%%%%%%%%%%%%%%%%%%%%%%%%%%%%%%%%%%%%%%%%%%%%%%%%%%%%%%%%%%%%%%%%%%%%%%%%%%%%%%%%%%%%%%%%
\section{Problem setting}
\label{setting}
We consider a non-trivial classification problem and will train single and multi-qubit variational quantum circuits to achieve this goal. The data is generated as a set of random points in a plane $x_{1},x_{2}$ and labelled as 1 (blue) or 0 (red) depending on whether they lie inside or outside of a given 2-dimensional non-convex figure. The goal is to train a quantum circuit to predict the label (red or blue) given an input point’s coordinate.
\section{Review of existing QML frameworks}
\label{Review of existing QML frameworks}
In this section we briefly review  three different necessary building block QML architectures for our software package  : hybrid-neural networks~\cite{schuld2020circuit,schuld2019quantum,farhi2018classification}, variational circuits~\cite{cerezo2020variational,bharti2021noisy} and data-reuplodaing~\cite{perez2020data,perez2021one}. 

%\subsection{Encoding of classical data into the quantum device}
%label{Dataencoding}

\subsection{Hybrid classical-quantum classifier (Hybrid)}
\label{hybrid}

Hybrid neural networks are formed by concatenating classical and quantum neural networks and can bring a great advantage by having a number of features in the initial classical layers that exceeds the number of qubits in the quantum layer. Normally we assume that in each layer we have one qubit for each feature and a sequence of one and two-qubit gates acting on it.
%%%%%%%%%%%%%%%%%%%%%%%%%%%%%%%%%%%%%%%%%%%%%%%%%%%%%%%%%%%%%%%%%%%%%%%%%%%%%%%
\begin{figure}[tbp]
\centering
\includegraphics[width=8.5 cm]{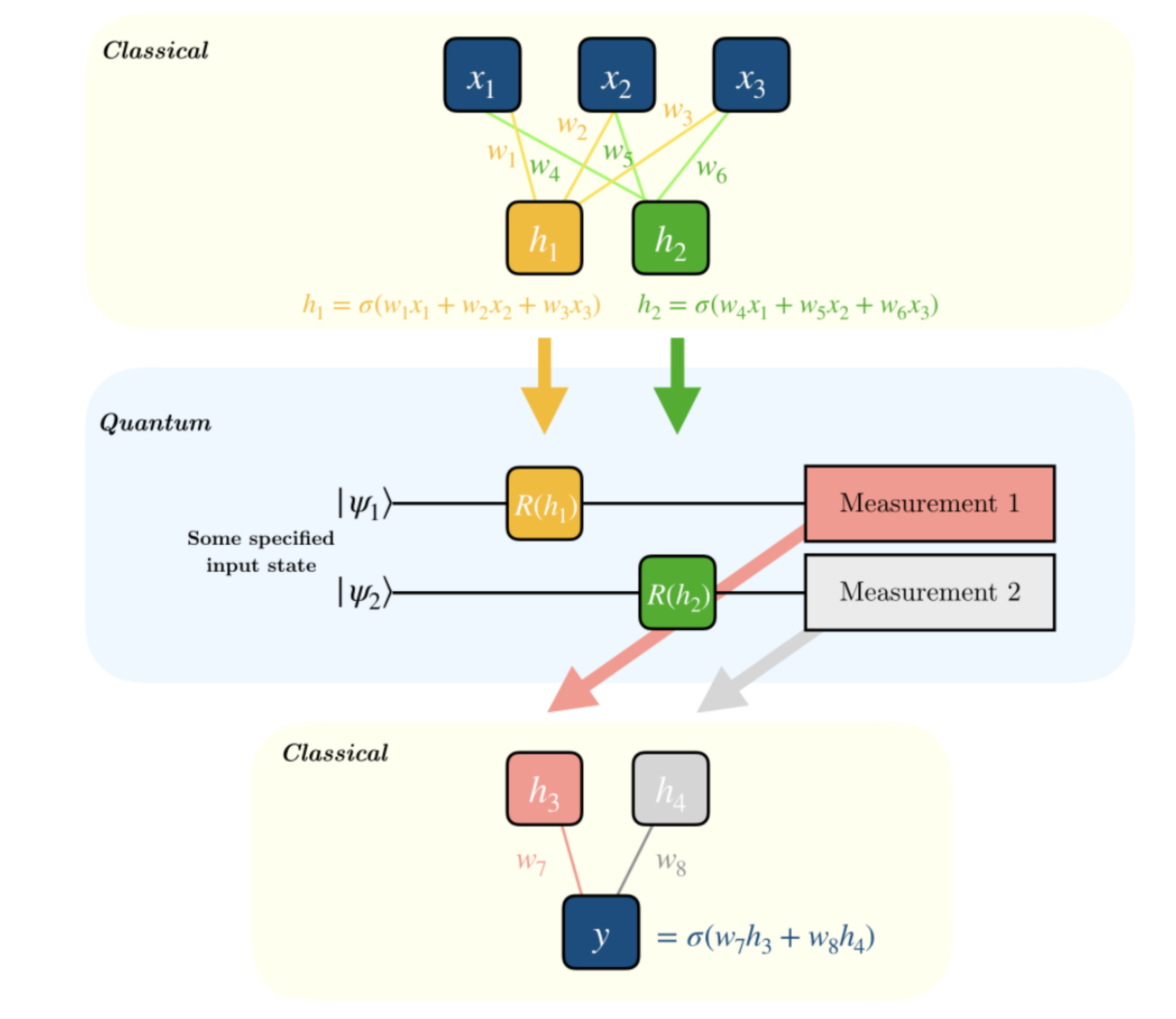}
\caption{Quantum circuit implementing hybrid classical-quantum classifier , each block corresponds to the layer of classical neural network.Taken from the Ref.~\cite{hybrid2020}.}
\label{reuploadscheme}
\end{figure}
%%%%%%%%%%%%%%%%%%%%%%%%%%%%%%%%%%%%%%%%%%%%%%%%%%%%%%%%%%%%%%%%%%%%%%%%%%%%%%%
To create a quantum-classical neural network a hidden layer is normally implemented utilising a parameterized quantum circuit. By "parameterized quantum circuit", we mean a quantum circuit where, for instance, the rotation angles for each gate are trainable parameters, specified by the components of a classical input vector. The outputs from our neural network's previous layer will be collected and used as the inputs for our parameterized circuit. Normally the measurement statistics of the quantum circuit can then be fed as inputs for the following layer. Notice that this kind of approach establishes a link between the classical and quantum neural networks.
An important point to note is that a single qubit classifier generates no entanglement, and can therefore be simulated classically. If one hopes to achieve a quantum advantage using hybrid neural networks, one needs to introduce several qubits and consequently entangle them, harnessing that quantum resource.
\subsection{Variational Quantum Algorithms(VQA)}
\label{VQE}
Variational circuits are quantum circuits that have learning parameters that are optimised through classical learning subroutines, in spirit, this kind of approach is reminiscent of a Variational Quantum Eigensolver (VQE)~\cite{cerezo2020variational,bharti2021noisy}.

%%%%%%%%%%%%%%%%%%%%%%%%%%%%%%%%%%%%%%%%%%%%%%%%%%%%%%%%%%%%%%%%%%%%%%%%%%%%%%%
\begin{figure}[tbp]
\centering
\includegraphics[width=9.0 cm]{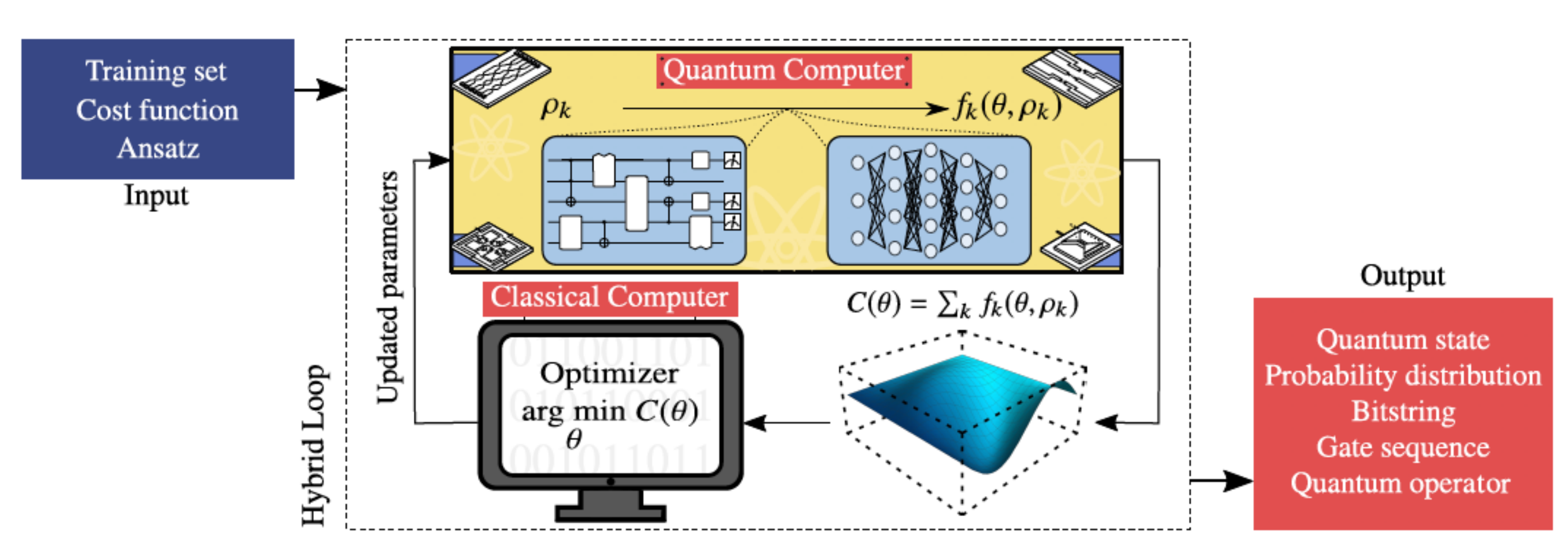}
\caption{Schematic diagram of a Variational Quantum Algorithm (VQA).Figure is borrowed from the Ref.~\cite{cerezo2020variational}.}
\label{VQA}
\end{figure}
%%%%%%%%%%%%%%%%%%%%%%%%%%%%%%%%%%%%%%%%%%%%%%%%%%%%%%%%%%%%%%%%%%%%%%%%%%%%%%%

As schematically shown in Fig.\ref{VQA}, the first step towards developing
a VQA is to define a cost or loss function C which encompasses the solution to the problem. After that, an ansatz is introduced through the quantum operation depending on a set of continuous or discrete parameters that can be optimized. This ansatz is then trained  in a hybrid quantum-classical loop to solve the optimization task at hand
\begin{equation}
\theta^{*}=\arg\min_{\theta} C(\theta).
\end{equation}

The trademark of VQAs is that a quantum computer is utilised to estimate the cost function $C(\theta)$ 
while harnessing the power of classical optimizers for training the quantum parameters. A rather crucial assumption here is that one cannot efficiently compute the cost function on the classical computer, as this would imply an absence of quantumm advantage in the VQA framework.

\subsection{DRC: Data-reuploading classifier}
\label{reupload}
Data re-uploading is a subclass of quantum embedding which is realised by catenating repeating units in a row. Single-qubit rotations applied several times along the circuit generate the necessary non-linearity for engineering a functional neural network. Moreover, it has been demonstrated that a single qubit can realise both being a universal quantum classifier~\cite{perez2020data} and being a universal approximant~\cite{perez2021one}
%%%%%%%%%%%%%%%%%%%%%%%%%%%%%%%%%%%%%%%%%%%%%%%%%%%%%%%%%%%%%%%%%%%%%%%%%%%%%%%
\begin{figure}[tbp]
\centering
\includegraphics[width=8.5 cm]{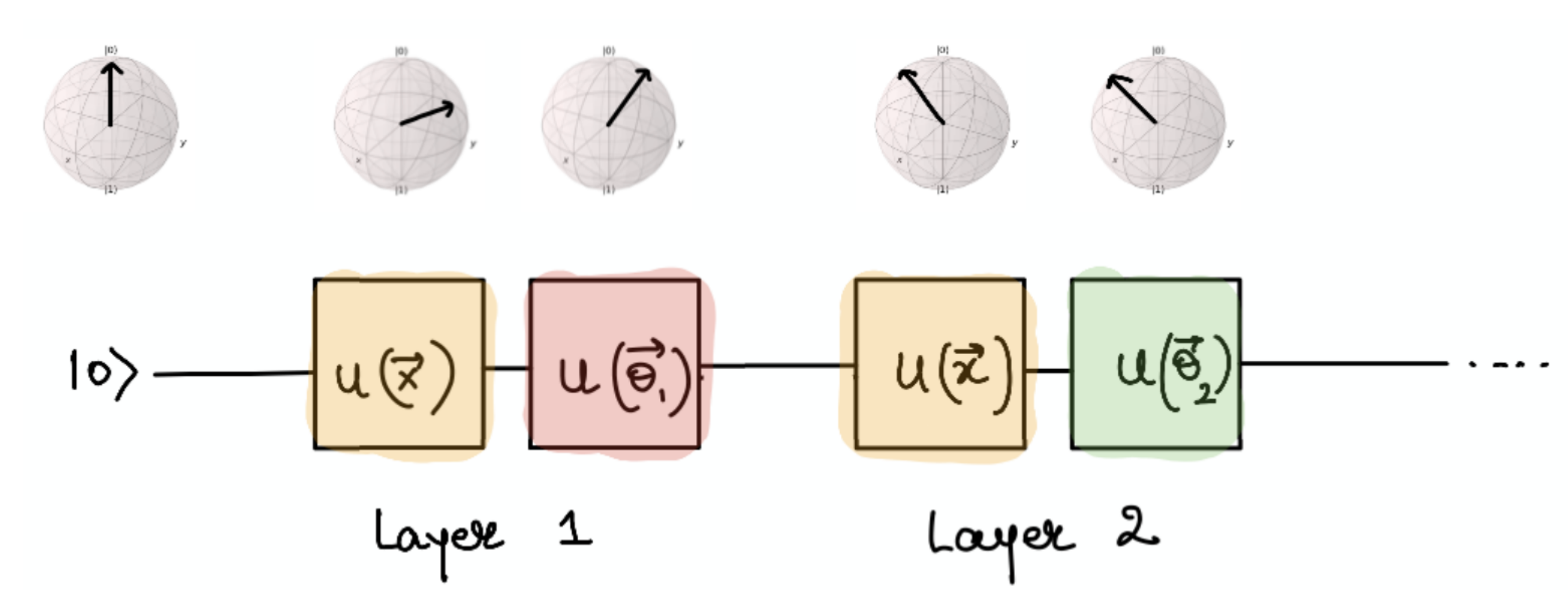}
\caption{Quantum circuit implementing data re-uploading, each block corresponds to the layer of classical neural network. Image is take from the Ref.~\cite{reupload2021}.}

\label{reuploadscheme1}
\end{figure}
%%%%%%%%%%%%%%%%%%%%%%%%%%%%%%%%%%%%%%%%%%%%%%%%%%%%%%%%%%%%%%%%%%%%%%%%%%%%%%%
To load $[x_{1},x_{2}]$ into the qubit, we just start from some initial state vector, $|0 \rangle$, apply the unitary operation $U(x_{1},x_{2},0)$ and end up at a new point on the Bloch sphere. Here we have padded 0 since our data is only 2-dimensional. Authors of Ref.~\cite{perez2020data} discuss how to load a higher dimensional data point $[x_{1},x_{2},x_{3},x_{4},x_{5},x_{6}]$ by breaking it down into sets of three parameters $(U(x_{1},x_{2},x_{3},U(x_{4},x_{5},x_{6})$.

\begin{figure}[tbp]
\centering
\includegraphics[width=8.5 cm]{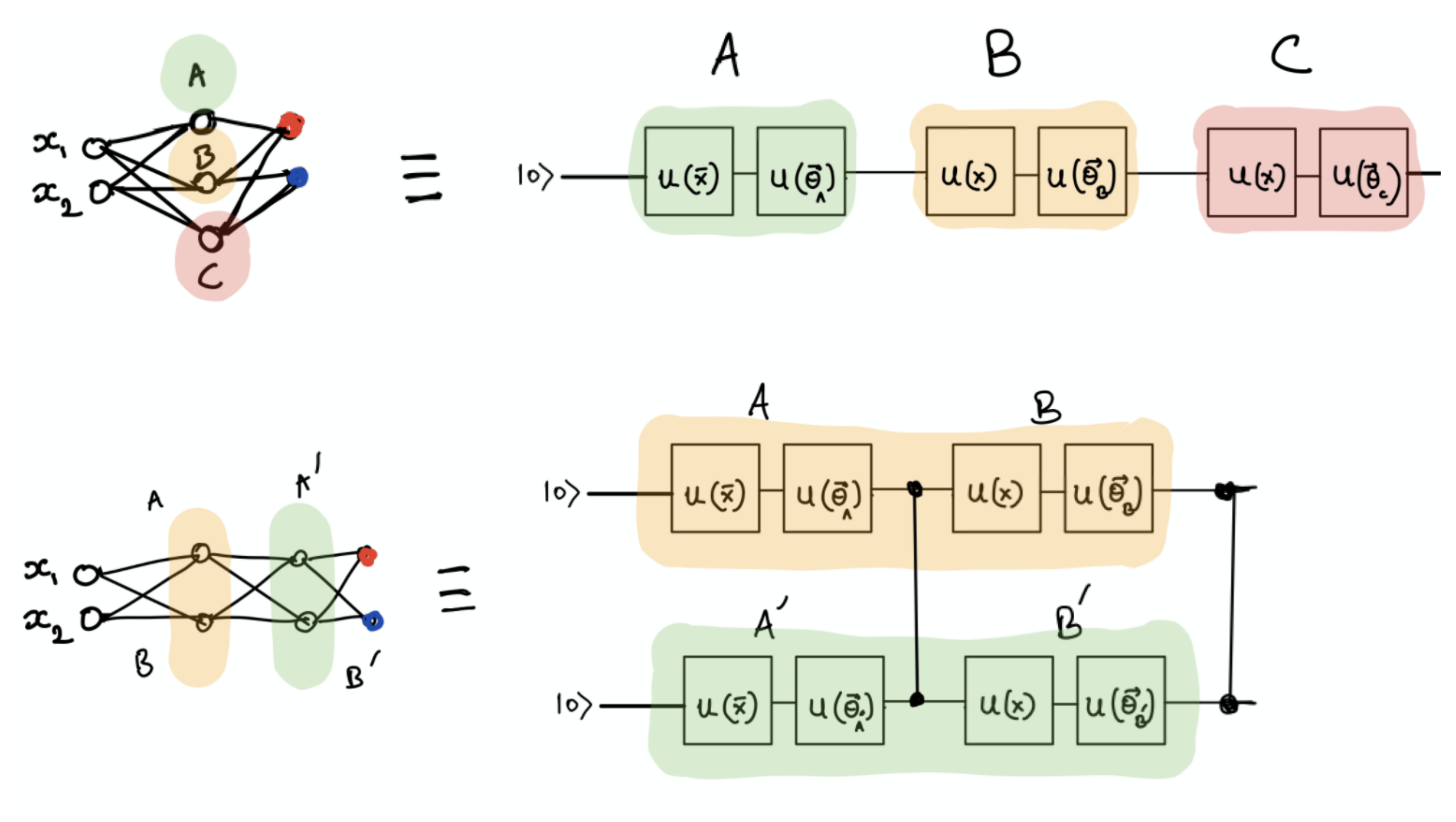}
\caption{Quantum circuit implementing data-reuploading with strongly entangling layers, where entanglement between the blocks  is introduced with controlled two qubit gates. Figure is take from the Ref.~\cite{reupload2021}.}.

\label{reuploadschemeentangle}
\end{figure}

After the data loading stage, we want to have some trainable non-linear model analogous to a deep neural network with a non-linear activation function where one can learn the weights of the model. Fig.~\ref{reuploadscheme1} are showing how data reuploading is implemented by the sequence of $B $ repeating units which correspond to the layers of classical neural networks, consequently one expects that with increasing  $B $ one gets a deeper neural network and consequently better learning can be obtained. Each unit is realised as a product of two unitaries $U(x_{1},x_{2},0)$ and $U(\theta_{1},\theta_{2},\theta_{3})$, where the second unitary contains the trainable parameters. This approach can be boosted by introducing strongly entangling layers through utilisation of CNOT gates as it is shown on  Fig.~\ref{reuploadschemeentangle}.

As it has been mentioned in the previous section, one can also speculate that multiple qubits with an entanglement between them could provide some quantum advantage over classical neural networks.
%%%%%%%%%%%%%%%%%%%%%%%%%%%%%%%%%%%%%%%%%%%%%%%%%%%%%%%%%%%%%%%%%%%%%%%%%%%%%%%
%%%%%%%%%%%%%%%%%%%%%%%%%%%%%%%%%%%%%%%%%%%%%%%%%%%%%%%%%%%%%%%%%%%%%%%%%%%%%%%

\begin{figure}[tbp]
\centering
\includegraphics[width=8.4 cm,height=4.0cm]{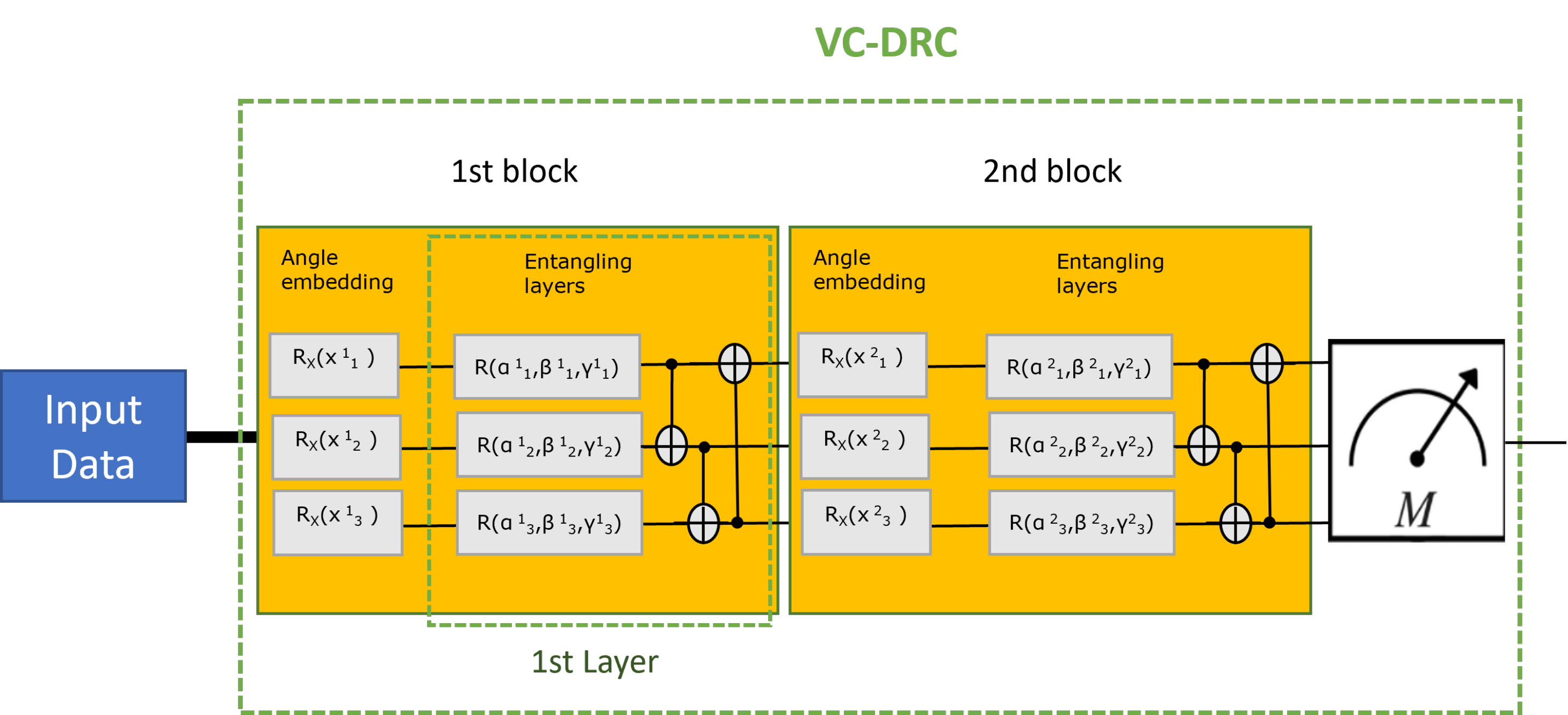} 
\caption{VC-DRC example for two Blocks and one entangling layer per block (B=2 , L=1).}
\label{Blockdiagram0}
\end{figure}

\section{FH:NN/VC-DRC and FH:VC-DRC/NN Full hybrid neural networks enriched with Variational  and data-reuploading technics}
\label{FULLHYBRID}

\begin{figure}[tbp]
\centering
\includegraphics[width=8.5 cm,height=4.0cm]{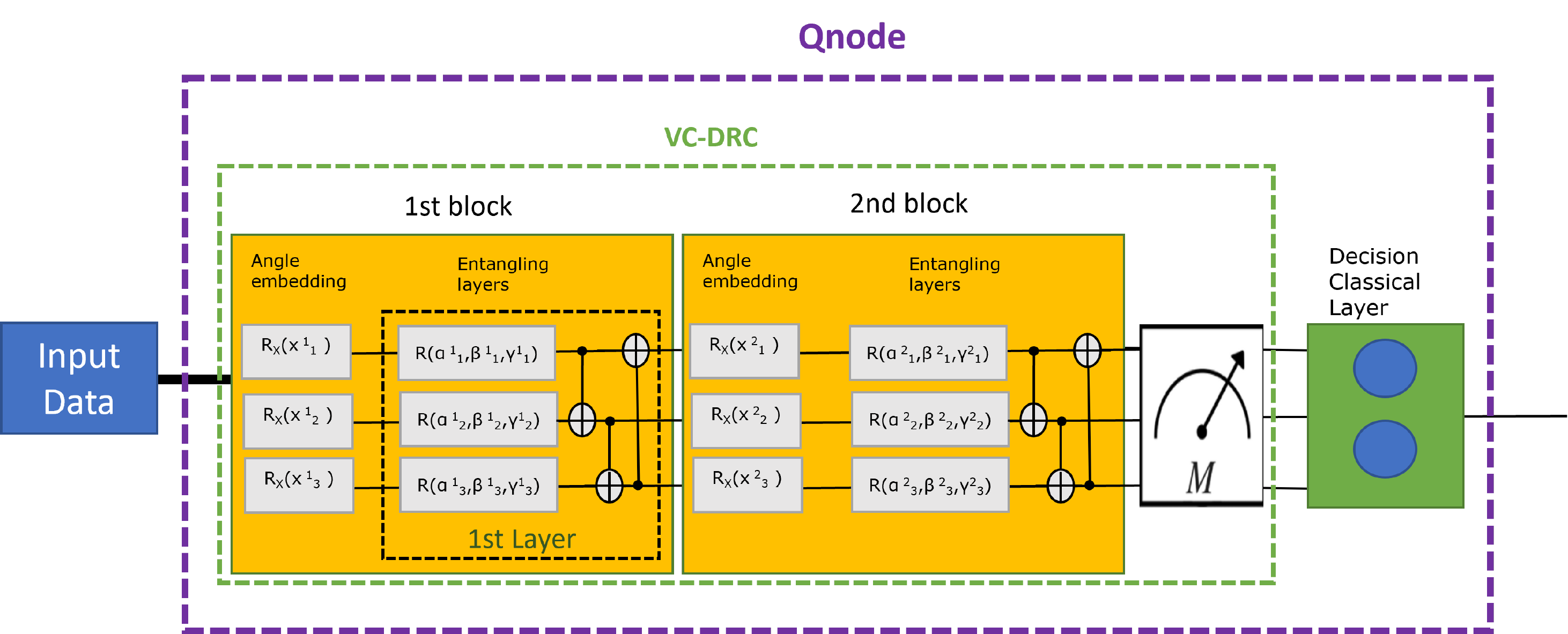}  \\
\includegraphics[width=8.5 cm,height=4.0cm]{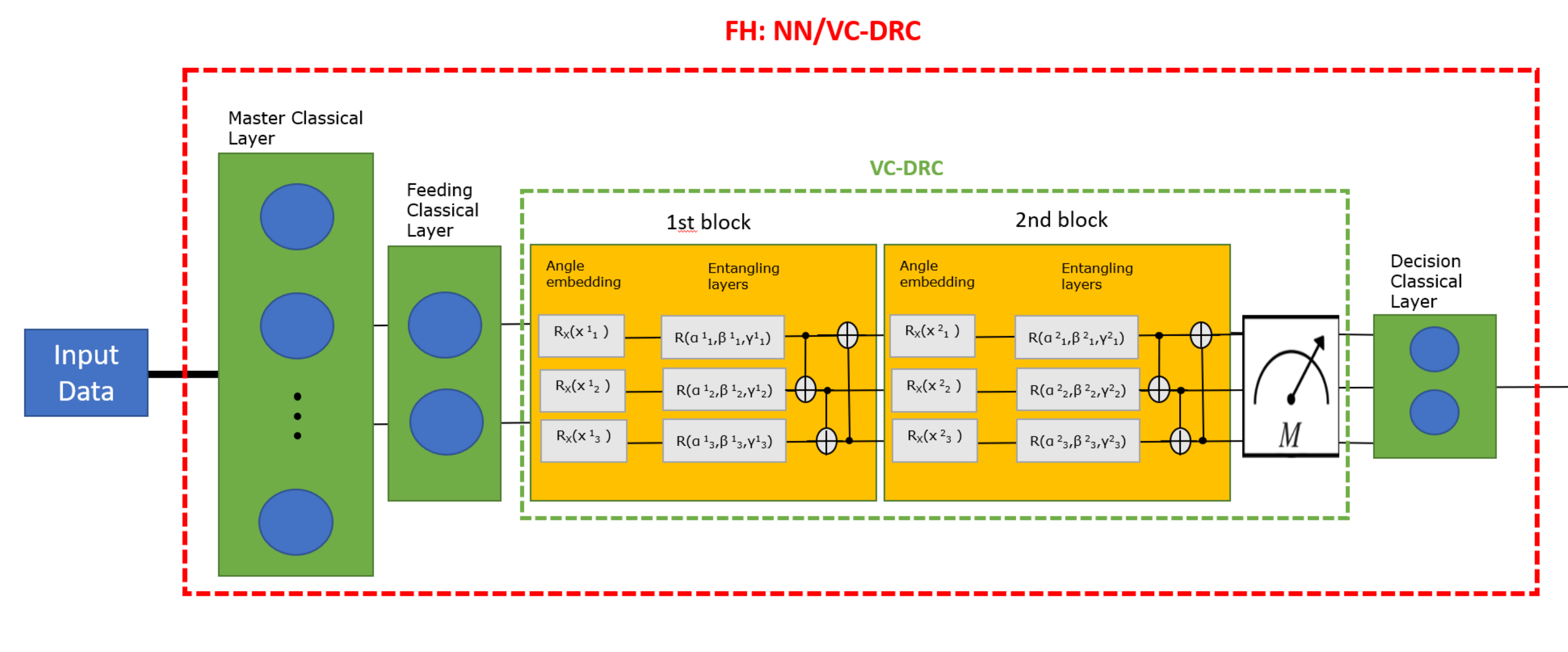} 
\includegraphics[width=8.5 cm,height=4.0cm]{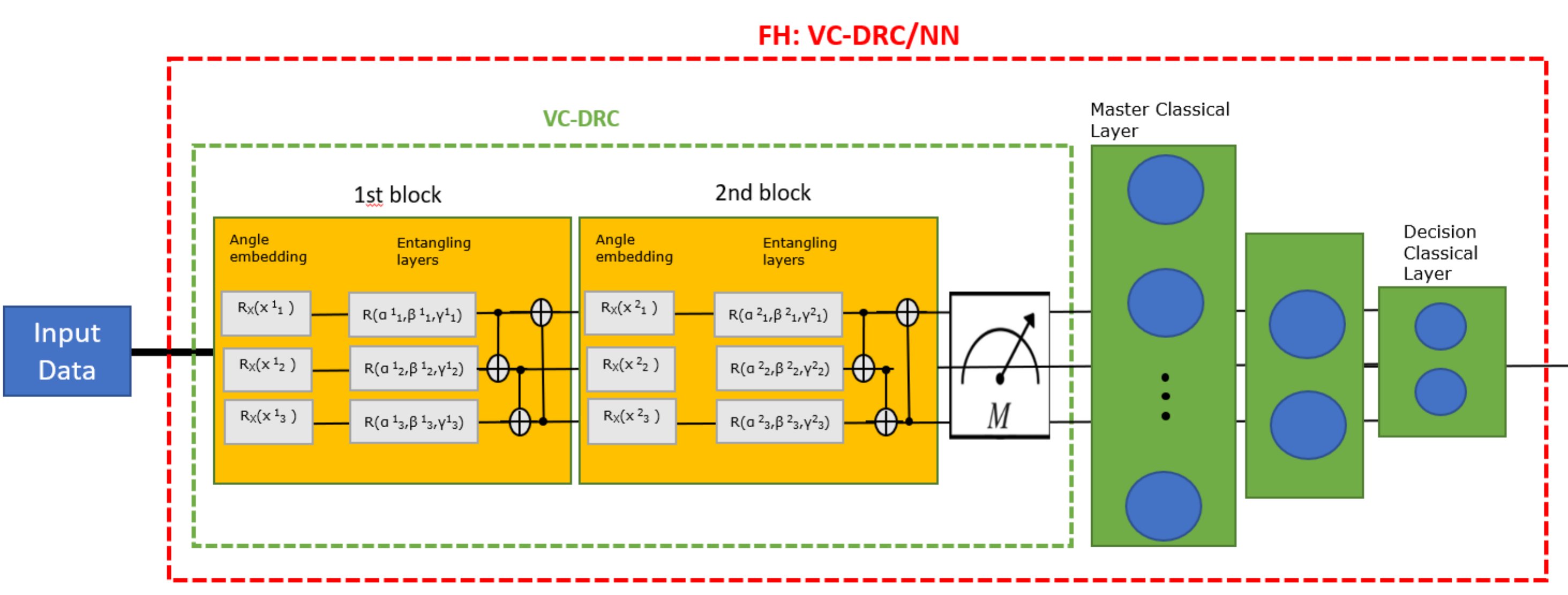}
\caption{(First row) Block Diagram of the specific QNode proposed in the current study. (Second row)  Block Diagram of the Full Hybrid (NN/VC-DRC) classifier where a VC-DRC circuit is placed after a classical neural network, (Third row)  Block Diagram of the Full Hybrid classifier where a fully quantum VC-DRC circuit is placed before classical neural network (VC-DRC/NN).}
\label{Blockdiagram}
\end{figure}
\subsection{VC-DRC}

 A Variational classifier circuit (VC) consists of a data embedding layer which in turn loads the classical data into the qubits followed by the entangling layers (CNOT gates that entangle each qubit with its neighbour) and the measurement outcome is the expectation value of a Pauli observable for each qubit~\cite{pennylane_variational}. In our case we use an angle embedding $R_x$.
In order to combine a VC circuit with DRC technique we define as one block (B) a sequence of data embedding and entangling layers (L). By adding many blocks we re-introduce the input data into the model.
In Fig.~\ref{Blockdiagram0} we illustrate such VC-DRC circuit for B=2 , L=1.

\subsection{QNode} 
Pennylane   is   an   open-source   software   framework   for   differentiable programming of quantum computers.  All our models are builded using this framework.
In Pennylane an object QNode represents a quantum node in the hybrid computational graph. Here a quantum function is used to create a quantum node, or QNode object, encapsulating the quantum function (corresponding to a variational circuit) and the device used to execute the function. 

Here we would like to clarify what we call a QNode in the scope of the current manuscript. As depicted in the first row of the Fig.~\ref{Blockdiagram}, a QNode is a specific circuit where input data are passed to the quantum Node which consists of a VC-DRC and a final classical decision layer.

The input classical data is passed into the quantum circuit as rotation angles $R_x$ ("angle embedding") on the Bloch sphere. After the computation on the quantum node is completed, measurement is performed and the outcome is passed to the classical decision layer which decides the final prediction label of the binary classifier.

\subsection{FH:NN/VC-DRC and FH:VC-DRC/NN} 

In this section we propose two varieties of a new binary classifier architectures which which are named under the common name Full Hybrid (FH). On a basic level, FH consists of a VC-DRC combined with classical layers.  We came up with two novel architectures for the FH circuits depending on whether the VC-DRC circuit is at the end or at the beggining of the model(named FH: NN/VC-DRC and FH: VC-DRC/NN respectively). These architectures are extensively studied in the current manuscript as novel candidates for performing binary classification on noisy datasets (See second and third row of Fig.~\ref{Blockdiagram}). Moreover, we demonstrate in great detail, in the next section, that FH architectures outperforms several previously known quantum classifiers and performs equally well compared to classical counterparts. We comment that, in the FH:VC-DRC/NN case the power of the approach is given by the fact that the VC-DRC part can be acting as a quantum embedding as evidenced by Refs.~\cite{lloyd2020quantum,mitarai2018quantum}. In this case the goal is to derive the angle embedding for which the separation of the data labels is maximized in the Hilbert space.
In what follows, we provide more technical details on the Full Hybrid architectures with an emphasis on explaining and giving more details on the Master, Feeding and Decision classical layers which are depicted in the second row of the Fig.~\ref{Blockdiagram}.

For FH: NN/VD-DRC the first part is a classical neural network(NN), followed by the VC-DRC circuit and a final decision layer, which is just a single neuron layer with a sigmoid activation function.  We use a classical NN that is not fine-tuned for this specific classification task. Moreover, as it can be seen on Fig.~\ref{Blockdiagram}, the classical NN can contain an arbitrary number of  layers, and each layer can contain an arbitrary number of neurons but the last layer (Feeding classical layer) should always have the same number of neurons as number of qubits. In our 2D case the classical NN, consists of a 2-neuron layer with ReLU as the activation function (Master classical layer), followed by a 2-neuron layer with a Leaky ReLU activation function (Feeding classical layer).  

For FH: VC-DRC/NN the first part is a VC-DRC circuit, followed by the previously described NN network  and the same final decision layer. We remark that we also tried Sigmoid , tanh and general geometric functions, and the best performing activation functions were selected.
%The role of the activation function on the feeding layer is not very important for 2D/3D cases and results do not vary significantly.      
%A variation of the Hybrid named (Hybrid 2) has been also used. The difference with the Hybrid is that the VC-DRC is placed before the Classical layers as can be seen in Figure 6. 
%In this way the VC-DRC acts as a training quantum embedding layer (the objective is of maximally separating data classes in Hilbert space thus acting as a quantum feature map: cite: Quantum embeddings for machine learning https://arxiv.org/pdf/2001.03622.pdf  --If one can find a faithful embedding, the computations required to compare data vectors and assign them to clusters can be performed using standard linear algebraic techniques.

\section{Comparative study of different quantum and classical classifiers}

\label{Comparative study of different frameworks}
Here we  test several models (including our proposed models) and benchmark them against each other as well as to the best-known classical machine learning counterparts by running on the simulator backends(such as Aer in qiskit) for  2-dimensional and 3-dimensional non-convex datasets. Then we will study the resilience of our different approaches to the noise by introducing asymmetrical Gaussian noise by studying the prediction grids and ROC/AUC characteristics. 

This kind of study sheds light on the learning properties as a function of the amount of existing noise in the dataset. These results have been obtained by systematic hyperparameter tuning, by observing how the ROC/AUC changes with: the number of repeating units in the data-re- uploading approach, batch size, number of epochs and the number of strongly entangling units. 
%Our study  provides an insight into the role of
%entanglement in boosting learning capabilities in QML frameworks.
%%%%%%%%%%%%%%%%%%%%%%%%%%%%%%%%%%%%%%%%%%%%%%%%%%%%%%%%%%%%%%%%%%%%%%%%%%%%%%%
\begin{figure}[tbp]
\centering
\includegraphics[width=3.0 cm]{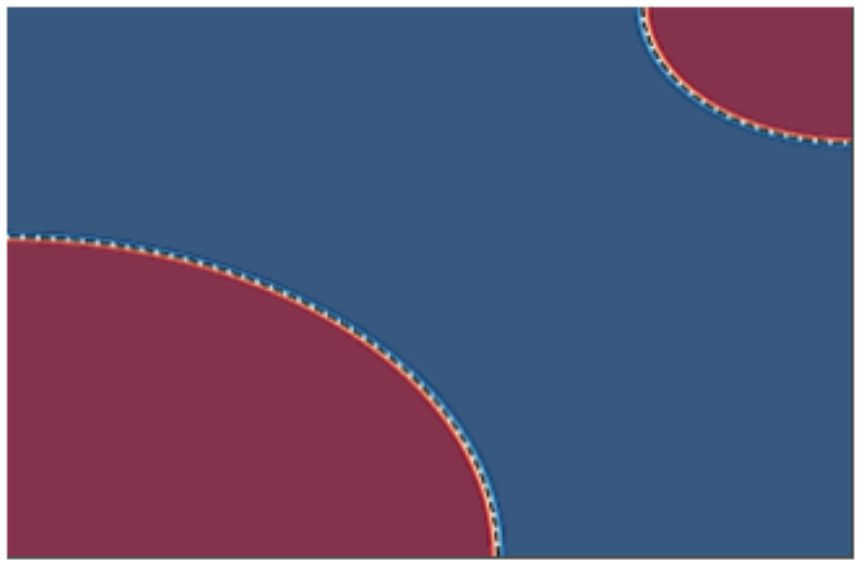} \\
\includegraphics[width=8.5cm]{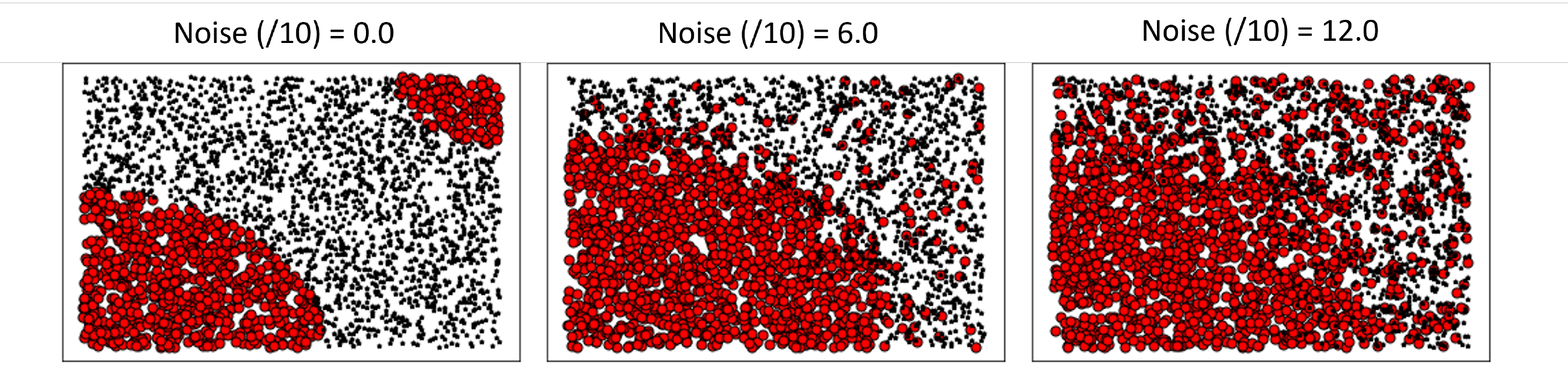}
\caption{(First row) Pattern of Dataset. (Second row) Pattern of dataset for different noise levels.}
\label{2data}
\end{figure}
To produce datasets with noise, we introduce asymmetrical (here noise is only applied to one class) Gaussian noise (N). In bottom Figure~\ref{2data} we plot the case of N=0.0 , N=0.6 and N=1.2. Each dataset has 6000 data points and is further equally splitted into training and testing datasets.

%\textbf{(Need to describe what noise level means),NIKOS, Also need to explain more on activation functions and most importantly loss function,maybe a give a formula for it}

%Single qubit Data reupload classifier  (DRC) [23, Data-reuploading classifier — PennyLane August 2021]
%2Q Variational classifier pennylane  (VC) [27, Variational classifier — PennyLane August 2021]
%2Q Variational classifier with data reupload (VC-DRC). Blocks and layers can be seen in the following schematic. For Blocks = 1 VC-DRC reduces to VC. For the following section we keep number of layers = 1 (except stated otherwise).
%It is important to outline that in the current article our novel approach is to use the DRC technique combined with a VC
%in a single model and in the quantum part of the hybrid classifier. This novel
%combination is expected to provide more robustness to dataset noise.This part is great it should be expanded
%Hyperparameters used for quantum models: Training epochs=30, Optimizer: Adam, Learning rate=0.02 Blocks=6 (unless stated otherwise), batch size = 16.
%Results:
%commenting on Figure 7

\begin{figure}[tbp]
\centering
\includegraphics[width=8.4 cm]{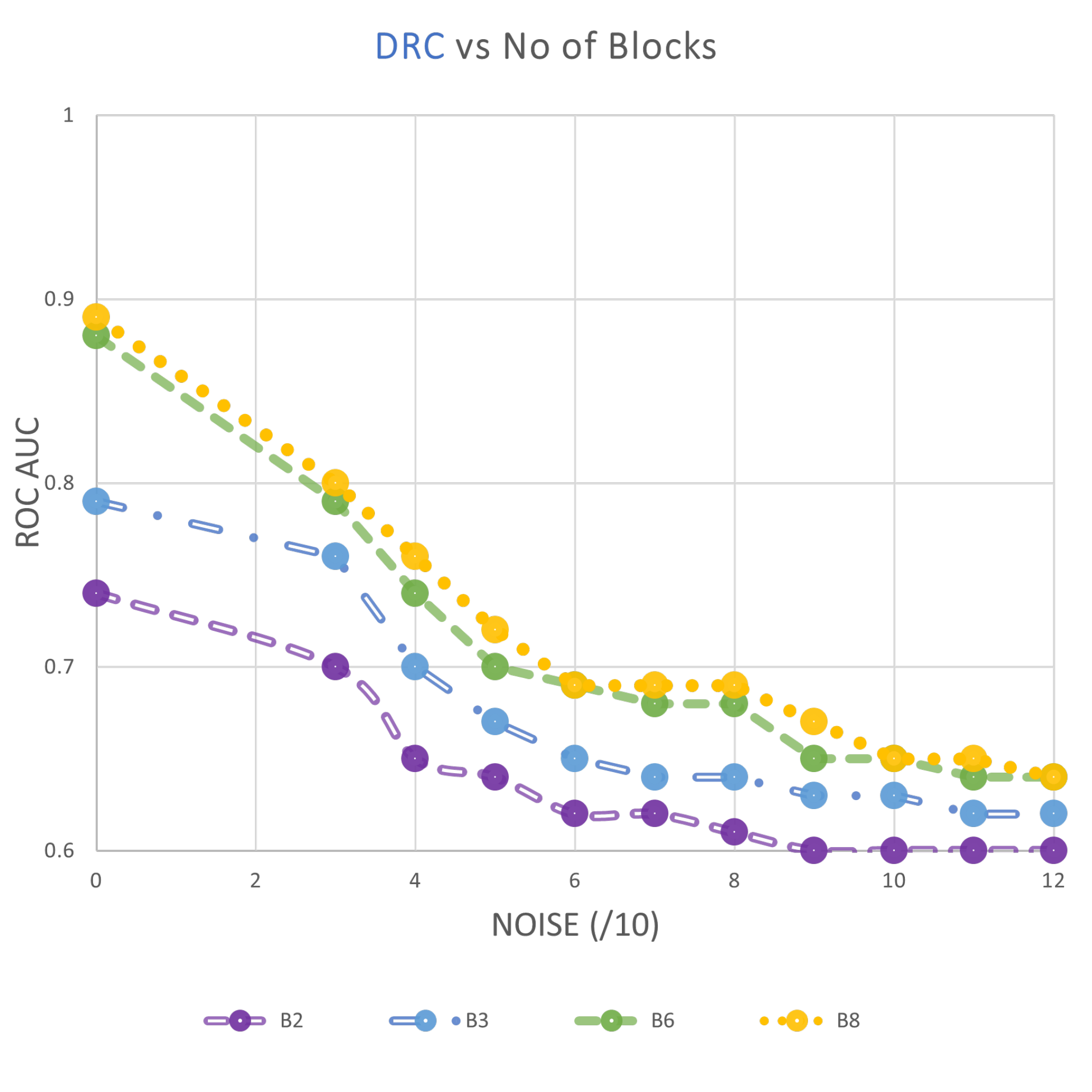} 
\includegraphics[width=8.4 cm]{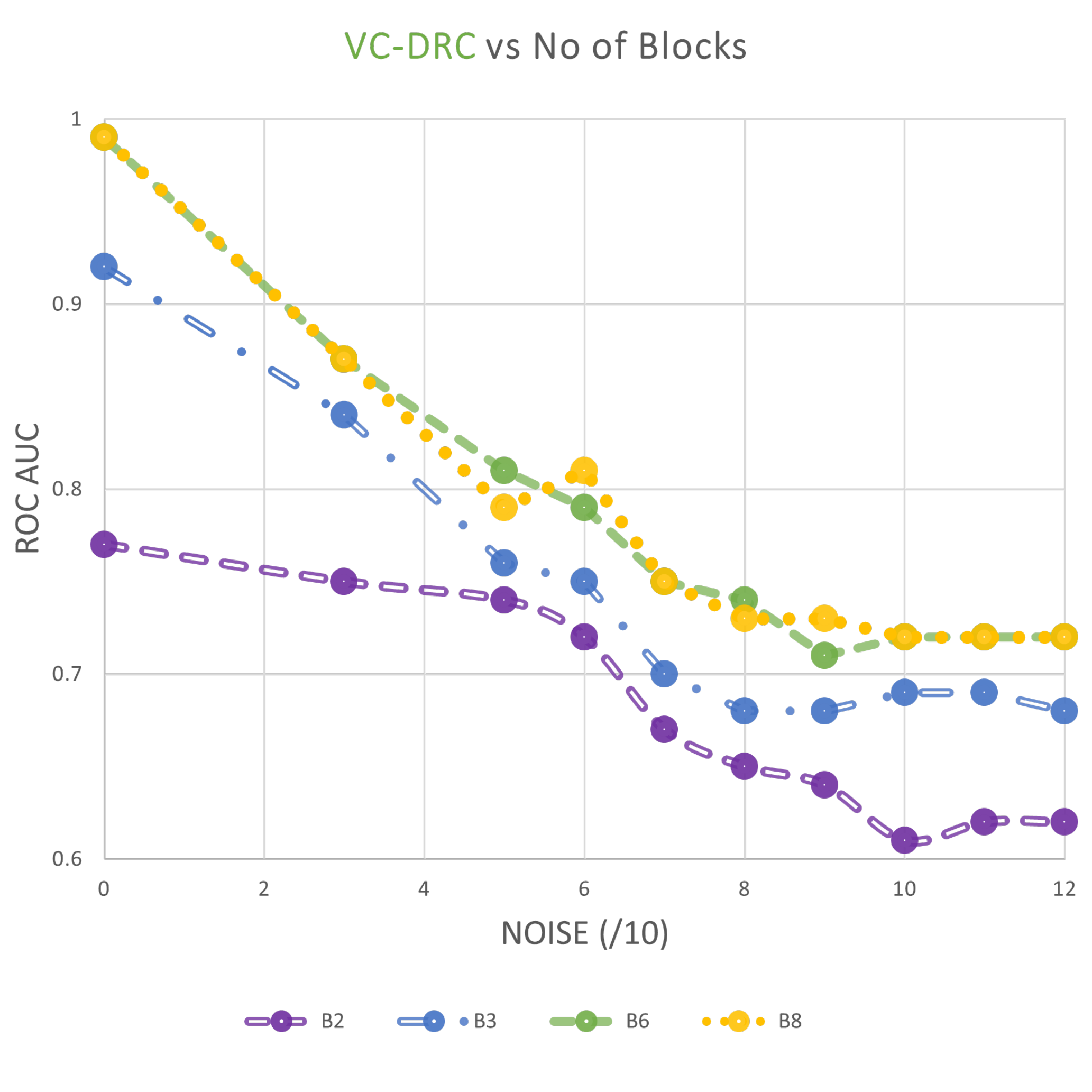} 
\caption{ROC/AUC as a function of number of repeating blocks of data re-uploading for the  DRC classifier (Top row) and VC-DRC classifier (Bottom row)}.
\label{Versusnumberofblocks}
\end{figure}

\begin{figure}[tbp]
\centering
\includegraphics[width=8.4 cm]{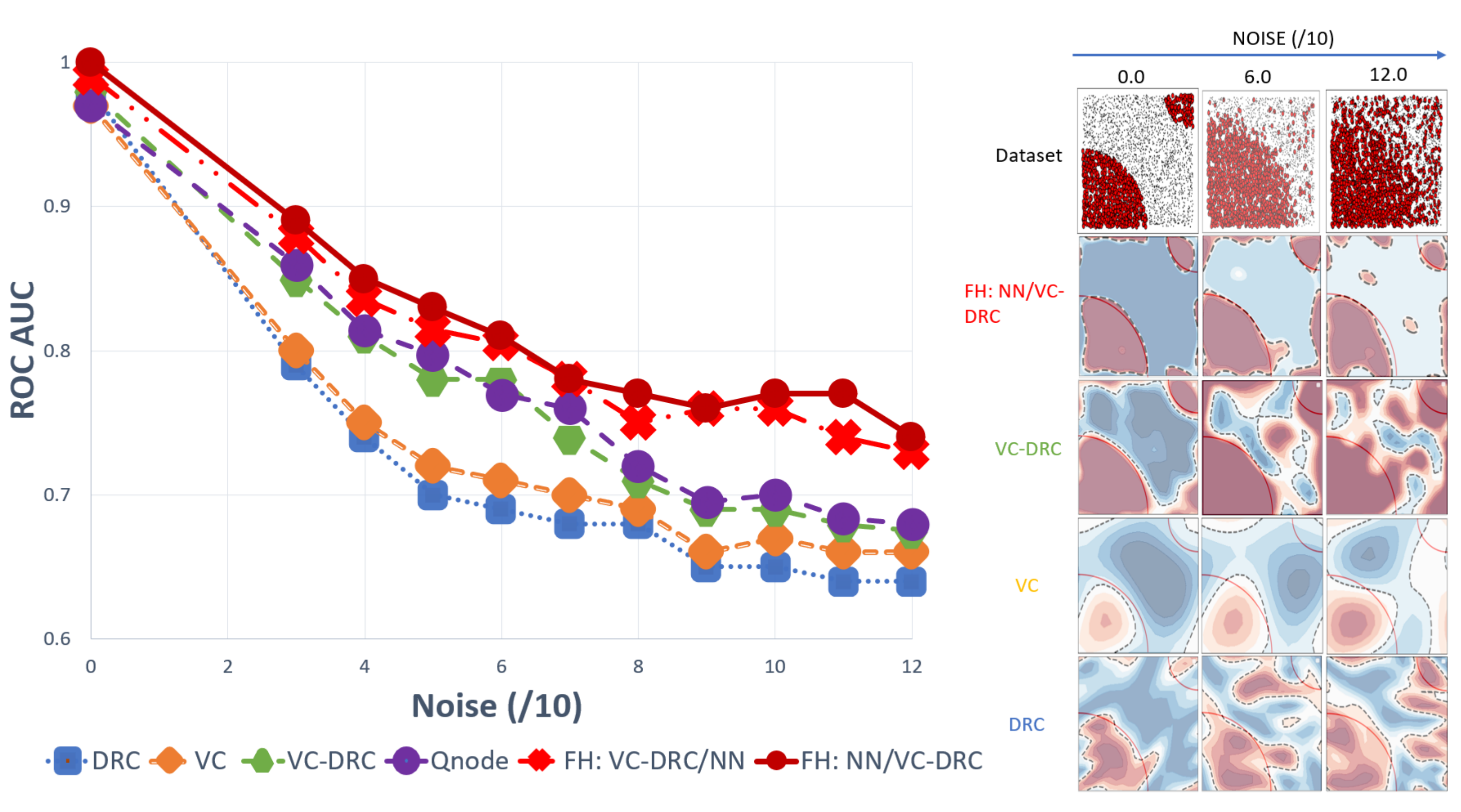} 
\caption{(Left) ROC/AUC for DRC, VC and VC-DRC classifiers for different noise levels. (Right) Prediction grids for respective classifiers for different noise levels.}.

\label{PredictionGrid2d}
\end{figure}

\begin{figure}[tbp]
\centering
\includegraphics[width=8.4 cm]{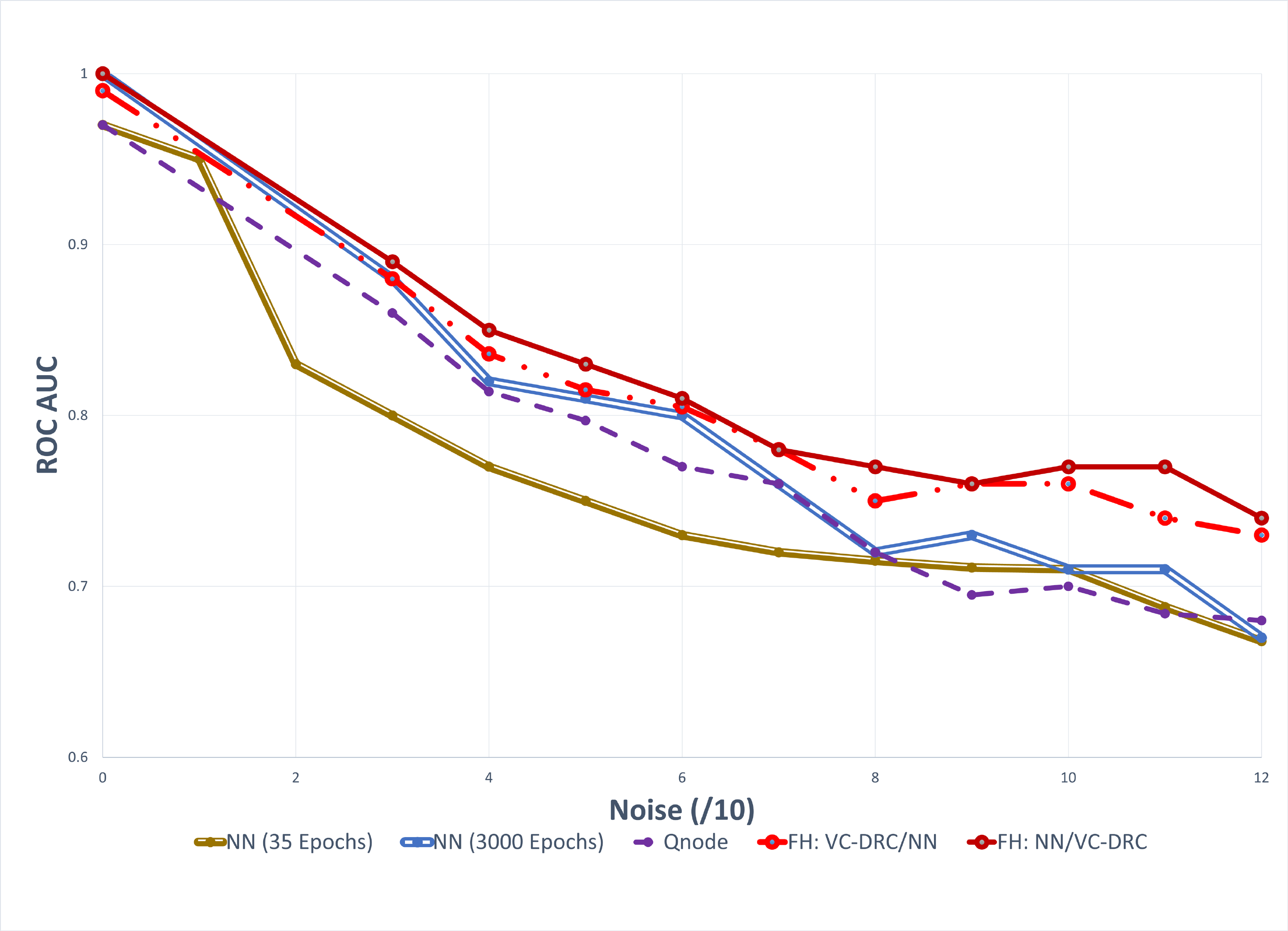} 
\caption{ROC/AUC for increasing level of noise for the classification of 2d dataset. Here we benchmark FH versus just the classical part (NN) and versus just the quantum part (QNode).}
\label{Benchmark1}
\end{figure}

To demonstrate the power of the data-reuploading technique combined with the variational classifier in the VC-DRC model, we plot the ROC/AUC versus noise for different number of blocks. The results are shown in Fig.\ref{Versusnumberofblocks}. It is apparent from Fig.~\ref{Versusnumberofblocks} (first row) that with an increasing number of repeating blocks, we get better ROC/AUC for every noise level for the DRC classifier. On the bottom row of figure  Fig.~\ref{Versusnumberofblocks} we show results for the VC-DRC where compared to DRC we get even  higher ROC/AUC. We remark that no major improvements are seen for a Block number greater than six. From now on, in all codes of this section, we will set the number of blocks equal to six (B=6). In what follows we specify number of blocks and layers for each classifier: 
1. The single qubit DRC  (B=6)
2. 2 qubit VC (with 6 layers, L=6) 
3. VC-DRC (B=6,L=1) 
4. QNode (B=6,L=1)
5. FH: VC-DRC/NN (B=6,L=1)
6. FH: NN/VC-DRC (B=6,L=1).
All models have been trained for maximum 35 epochs, using the same optimizer and learning rate. The best result during the training process is shown.  
%commenting on Figure 8
On the left Fig.\ref{PredictionGrid2d} we compare all the previously mentioned classifiers. 
As we can see from on the left Fig.~\ref{PredictionGrid2d}  VC-DRC outperforms both VC and DRC.VC-DRC and Qnode have almost identical performance. The FH:NN/VC-DRC outperforms all classifiers whilst FH:VC-DRC/NN  has slightly worse behavior. 
In the right column of Fig.\ref{PredictionGrid2d} we can see the prediction grids for all classifiers at different noise levels.
For low noise levels (Noise/10=0), DRC and VC struggle to capture the prediction grid pattern while VC-DRC and FH almost capture it. For medium noise levels (Noise/10=6), DRC tends to capture the noise (overfitting) while VC looks more stable. VC-DRC still captures the main pattern but also shows signs of overfitting. FH performs very well thanks to the classical preprocessing and utilising the power brought by VC-DRC. For high noise levels (Noise/10=12) FH captures the pattern and shows robustness to the noise while the rest of the classifiers are capturing the noise.
%commenting on figure 9
In order to demonstrate that FULL HYBRID does not perform well only because of the strong classical NN attached to the quantum circuit,  we benchmark FH versus just the Classical part (NN) and versus just the Quantum part (QNode). On from Fig.~\ref{Benchmark1} we show results for two NN's one with 35 epochs training (same training epochs as in the FH) and 3000 epochs just to see what is the best outcome this NN can produce. We conclude that the FH outperforms both it's components (NN and QNode) which shows that FH is more powerful classifier than it's isolated parts. 

\begin{figure}[tbp]
\centering
\includegraphics[width=8.5 cm]{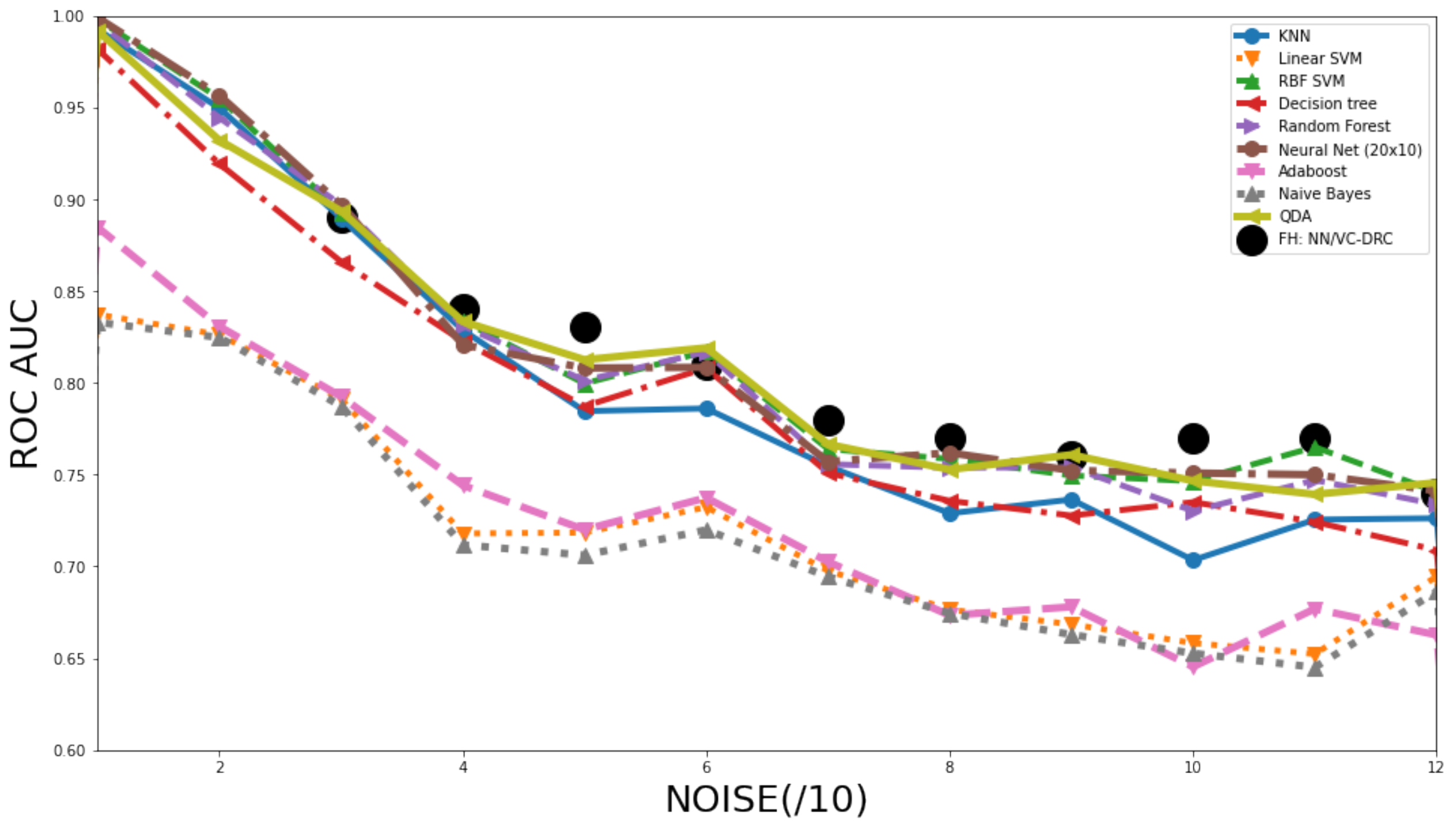} 
\caption{ROC/AUC for increasing level of noise for the classification of 2d dataset. Here we benchmark a great number of classical classifiers against our proposed FULL HYBRID classifier.}
\label{Benchmarkclassical}
\end{figure}

\begin{figure}[tbp]
\centering
\includegraphics[width=8.5 cm]{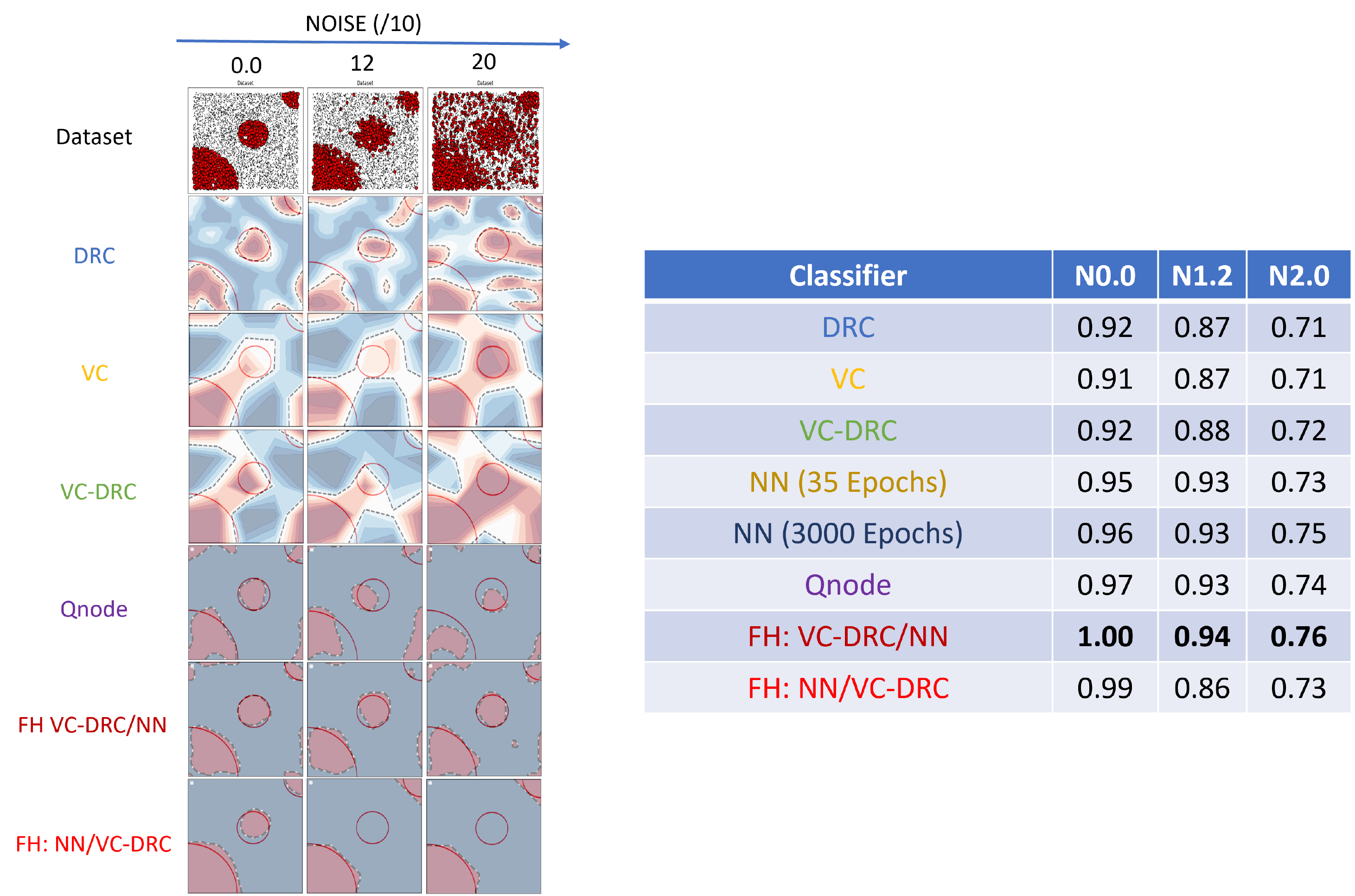} 
\caption{(Right) ROC/AUC for  classifiers for different noise levels. (Left) Prediction grids for respective classifiers for different noise levels.}

\label{tripledataset}
\end{figure}

%commenting on figure 10
To test even further the FH classifier, we benchmark its performance against a great number of classical counterparts, which are specified in the inset of the  Fig.\ref{Benchmarkclassical}. Interestingly, this figure shows that in the high noise region, the quantum classifier even outperforms some classical ones or performing equally well in all noise regions. We also see that compared to the other classical approaches (QDA, Decision tree, KNN and Random forest) that are well suited for non-convex classification problems and showing good performance in all noise regimes.
%commenting on figure 11
In Fig.~\ref{tripledataset} we are showing results for a more complicated non-convex classification problem versus noise. In the table on the right we summarize the highest ROC/AUC scores for the respective classifiers. In the left figure we show prediction grids for the respective quantum classifiers. As in previous case, VC is more stable to noise and DRC tends to overfit and explores richer prediction grids. That is why VC-DRC, which combines both features, and the more complex approach like FH, is giving great results as apparent from row number 6. Surprisingly, for this particular dataset FH: NN/VC-DRC fails to capture the pattern of the dataset while FH: VC-DRC/NN captures the pattern and has the highest ROC/AUC score. It should be noted that the FH models outperforms again both it's components (NN and QNode). 

%VC for the binary classification is that it is more robust against noise but struggles to
%capture the patterns (features). DRC does not capture the main pattern in high level noise , moreover it tends to capture the noise which can be interpreted as overfitting.  VC-DRC captures the main pattern and some noise.  Full Hybrid model captures the pattern and shows robustness to the noise. }

%\begin{figure}[tbp]
%\centering
%\includegraphics[width=7.5 cm]{Latex_File_21_July/ROCAUCNOISE.png} 
%\caption{ ROC/AUC for the QNODE(purple dots) , Hybrid(brown dots),Hybrid Qnode first (red) %MLPC (grey dots) classifiers for different noise levels.}.

%\label{ROCAUC2D}
%\end{figure}

%The five previously mentioned models and a classical model (k-nearest neighbors (MLPC)) have been benchmarked
%using a 2D artificial datasets. A two class (red and blue) dataset was artificially
%generated (Fig. 5 left). In order to mimic a noisy unbalanced dataset, we introduced
%asymmetrical Gaussian noise and the AUCROC metric was evaluated for each
%model for increasing levels of noise. The results (Fig. ~\ref{toughcase}) the hybrid model was
%second-best the classical KNN model overall and the DRC model performed equally
%well with KNN at low noise levels. Corresponding prediction grids for VC-DRC and
%the hybrid model versus Asymmetrical Gaussian noise (Fig~\ref{comparison}) reveal that the VC-
%DRC model captures the pattern and is robust against noise while the hybrid model
%offers a different and richer prediction grid.  

% \knote{$\mathbb{U}$ is a unitary, while $\mathbb{U}_K$, $\mathbb{CS}$ are states. This is a bit unclear notation... maybe remove the naming of $\mathbb{U}$, they do not need an explicit name, e.g.  $\left\{ U_{i}\right\}_{i=1}^{r}$ instead of $\mathbb{U}\equiv\left\{ U_{i}\right\}_{i=1}^{r}$
% We should define $U_K$ here.}
% \end{defn}

% \begin{defn}
% Given a set of unitaries $\mathbb{U}\equiv\left\{ U_{i}\right\} _{i=1}^{r}$,
% a non-negative integer $K$ and some quantum state $\vert\psi\rangle,$ let
% $\mathbb{S}_{K}$ be the set of $K$-moment states. 
% \end{defn}

\medskip
%\noindent {\em Examples.---}

\section{Conclusions and Future Directions}
\label{conclusions}

%1.briefly mention what we did in this paper
%what are the main findings and why they are important?
%what are the communities that our findings are relevant for??
%Summary needs to mention about prediction greeds
%honest opinion on how we perform compared to the best classical NN classifiers
In this paper, we applied Quantum Machine Learning frameworks to improve binary classification models for noisy datasets which are prevalent in financial markets. The metric used for assessing the performance of our quantum classifiers is the area under the receiver operating characteristic curve (ROC/AUC). By combining such approaches as hybrid-neural networks, parametric circuits, and data re-uploading we created a new approach called Full Hybrid (FH). We tested our models for the classification of 2 and 3-dimensional non-convex  datasets and benchmarked them against each other as well as to the best known classical machine learning counterpart by running simulations on quantum simulators. Then, by introducing asymmetrical Gaussian noise in the input datasets, we studied the resilience of our different approaches to noise. This kind of study sheds light on the learning efficacy to the amount of noise in the dataset. In the scope of the manuscript we also performed systematic hyperparameter tuning by studying how ROC/AUC changes with the number of repeating units in the data-re-uploading approach, number of qubits, batch size, number of epochs and number of strongly entangling units. 
%2.what are the main results,main findings
 An extensive benchmarking of our new QML approach against existing quantum and classical classifier models reveals that our novel (FH) models exhibits better learning properties to asymmetric Gaussian noise in the dataset compared to known quantum classifiers, and performs equally well for existing classical counterparts. Yet more understanding of the merits of the (FH) classifier has been gained by a detailed analysis and comparison of the prediction grids for the VC, DRC, VC-DRC, QNode binary classifiers. 
 We observed that for low noise levels , DRC and VC struggle to capture the prediction grid pattern while VC-DRC and FH almost fully capture it. For medium noise levels, DRC tends to capture the noise (overfitting) while VC looks more stable. VC-DRC still captures the main pattern but also shows signs of overfitting. FH performs very well thanks to the classical preprocessing and utilising the power brought by VC-DRC. For high noise levels, (FH) captures the pattern and shows robustness in noise while the rest of the classifiers are capturing the noise in the dataset. 
%future directions 
 %Barren Plateu
It is a well conceived fact that one of the bottlenecks for VQAs is the phenomenon called "barren plateau"~\cite{mcclean2018barren}. As it has been demonstrated in Ref.~\cite{mcclean2018barren}, a given spin-spin interacting Hamiltonians cost function may exhibit a barren plateau, associated with exponentially vanishing variance in its first derivative, when one increases the number of qubits. Moreover, the VQE based algorithms perform a classical-quantum feedback loop to update the parameters of the parametric quantum circuits. 
%Kishors papers mention on minimizing the feedback loop
For future studies, it would be interesting to implement non-VQA algorithms for building more efficient quantum classifiers. By the time a classical computer calculates its  output, the classical-quantum feedback loop limits the efficiency of the quantum device, slowing the algorithm execution on current cloud computing frameworks. Most of the obstacles faced by VQE, such as the barren plateau issue~\cite{mcclean2018barren} as well as lacking a systematic method to select the ansatz and the innate necessity of having controlled unitaries, have been recently tackled  by suggesting a quantum assisted simulator (QAS)  \cite{haug2020generalized,bharti2020quantum}. Remarkably, The QAS algorithm does not require any classical-quantum feedback loop, can be parallelized, alleviates the barren plateau problem by prescribing a systematic approach to constructing the ansatz, and is not based on the usage of complicated unitaries.
%1. study of quantum noise
Of course, for the future studies, one has to keep in mind that sensitivity to errors and noise in qubits and quantum gates are the two most prominent obstacles towards scalable universal quantum computers. Given that, it would be nice to study how our results are affected if one implements noise models for realistic quantum backends. In general, a noisy quantum system is described by the open system model and systems dynamics within the Born-Markov approximation is governed by the Lindblad master equation for the system's density matrix~\cite{carmichael1993master}. Another approach to describe the different noise channels is based on Kraus operators which are the most general physical operations acting on density matrices\cite{nielsen2002quantum}.
Since most of our codes were based on PennyLane, it is instructive to mention that Pennylane has 3 different ways for implementing  noise in quantum circuits: classical parametric randomness, PennyLane’s built-in default.mixed device, and plugins for other platforms. Of course, Quantum circuits may be run on a variety of backends, some of which have their own associated programming languages and simulators. PennyLane interfaces to these other languages via plugins such as for Cirq and Qiskit. 
%The Qiskit simulator Aer is provided by qiskit.aer, with noise operations living in qiskit.providers.aer.noise. Qiskit has a different approach, incorporating a noise model into the device itself. 
%3. applying to the real world financial market dataset running on real quantum backends
Finally, it is also worth mentioning that we plan to test our classifiers on real world financial data. Here we hope to demonstrate that our proposed classifiers have the potential to improve credit scoring accuracy. Credit scoring provides lenders and counterparties better transparency of the credit risk they are taking when dealing with a counterparty. For large companies, this transparency is provided by public credit ratings. Small and medium enterprise companies(SMEs) are not covered by rating agencies and are suffering from reduced availability of credit. These datasets, along with the best classical neural networks, will by provided by the company called Tradeteq (Tradeteq is a value-added service provider to the Networked Trading Platform (NTP) of Singapore).

In summary, we have demonstrated  that the FH architecture outperforms several previously known quantum classifiers along with some of the best known classical counterparts. Interestingly, in the FH: VC-DRC/NN case, the power of the approach is given by the fact that, the VC-DRC part is acting as quantum embedding.
%\appendix
%\clearpage

\bigskip

\appendix

\section{Benchmarking binary classiffier for 3 dimensional non-convex datasets}
\label{Appendix}
\begin{figure}[tbp]
\centering
\includegraphics[width=8.5 cm]{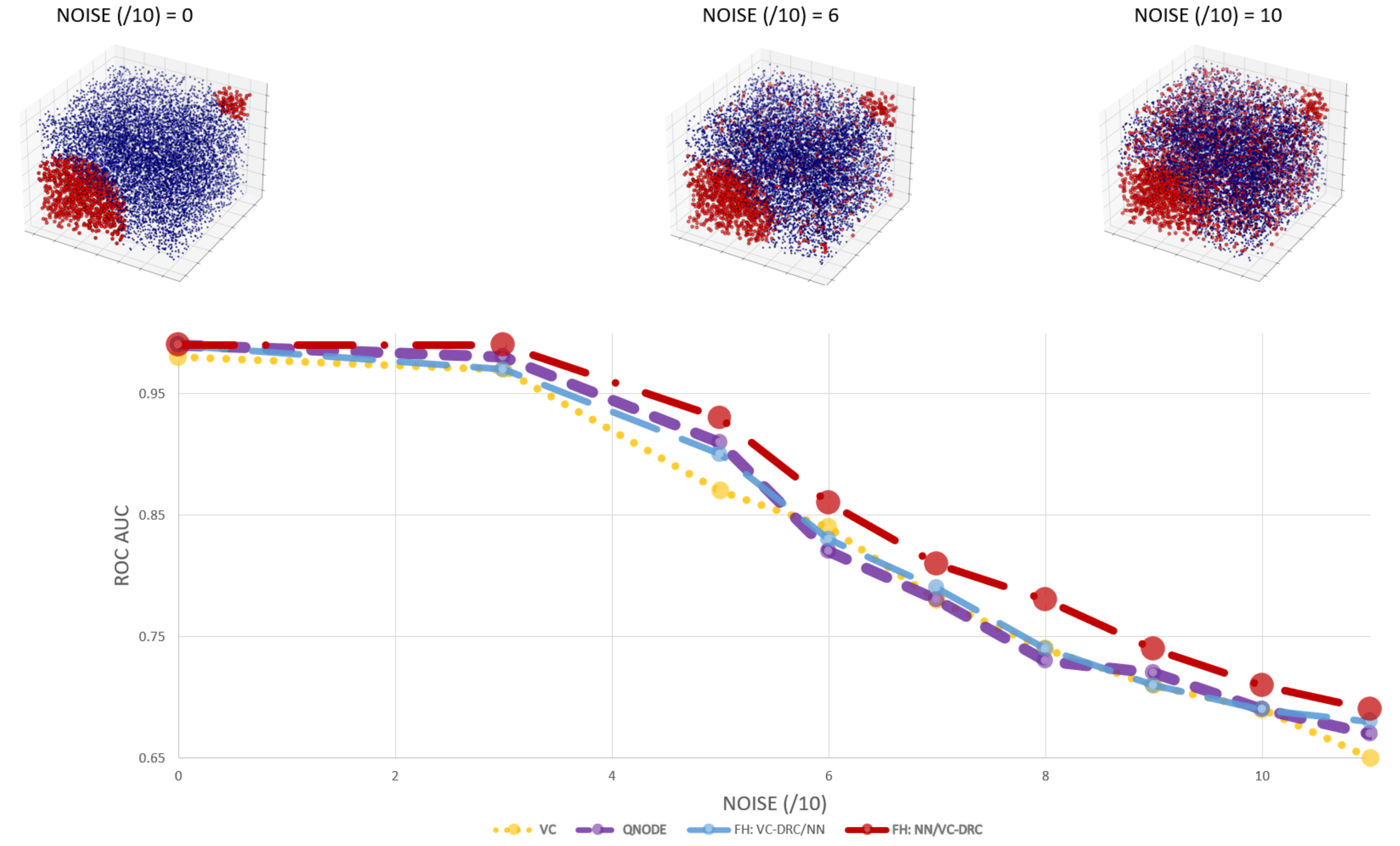} \\

\caption{(First Row) Dataset for the 3 dimensional non-convex problem, with increasing noise levels from left to the right (Second row) ROC/AUC for the QNode (red dots) , Full-Hybrid (grey dots) and MLPC (orange dots) classifiers for different noise levels.}

\label{Figure 10}
\end{figure}

\begin{figure}[tbp]
\centering
\includegraphics[width=8.5 cm]{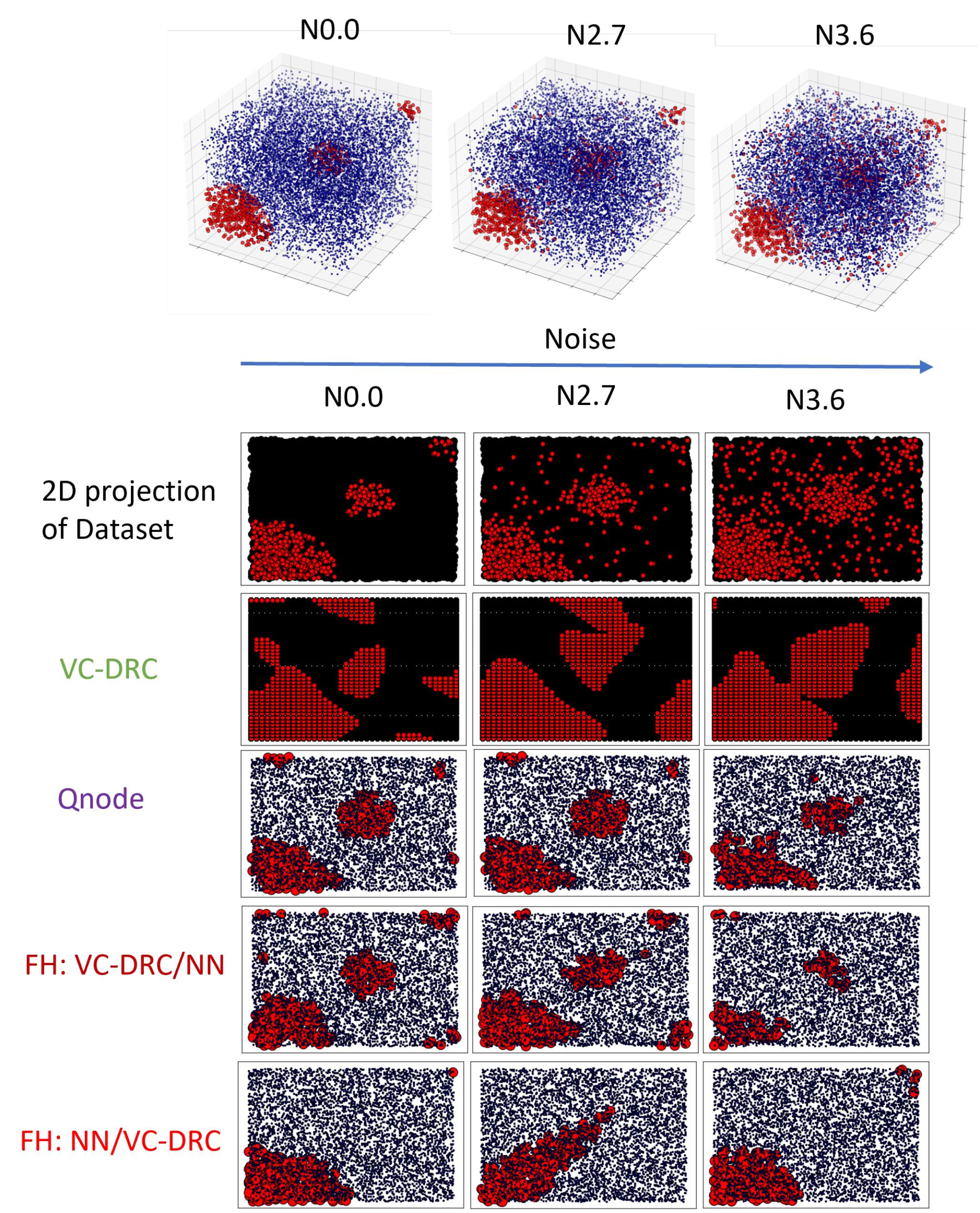} 
\caption{(First row) ROC/AUC for classifiers for different noise levels for 3D dataset. (Second row) 2D projection prediction grids for respective classifiers for different noise levels.}

\label{Figure 13}
\end{figure}

In order to gain insights about the performance of our FH classifier architecture in higher dimensions we study classification of 3-dimensional non-convex figure, dataset of which for the different levels of noise is presented in the first rows of Fig~\ref{Figure 10} and Fig~\ref{Figure 13}. Then we proceed by benchmarking FH classifiers against the QNode and VC. As we can see for this particular dataset, FH:NN/VC-DRC shows the best performance.
 
We also found it interesting to show (See Fig~\ref{Figure 13}) a 2D projection of the prediction grid for the dataset which has 3 red regions and thus is the hardest non-convex example considered so far. In rows 2-6 we show prediction grids for the VC-DRC, QNode, FH:VC-DRC and FH:NN/VC-DRC, respectively. We see that in the low noise regime FH:VC-DRC/NN and QNode perform the best at the same time, VC-DRC fails to capture upper right corner of that red region. Similar findings hold for the intermediate noise regime, However in the high noise regime FH:VC-DRC/NN and QNode obtain only two regions in red due to too much noise in the dataset, yet VC-DRC is still largely overfitting showing 3 disconnected regions, but guessing their locations wrongly. To summarise, FH classifiers show the maximal resilience to the noise in the dataset and are well suited for non-convex boundary classification problems.

%\begin{figure}[tbp]
%\centering
%\includegraphics[width=12.0 cm]{Latex_File_21_July/Figure 11 ST2 3D Classical comparison.png} 
%\caption{AUC/ROC for increasing level of noise for the classification of 3d dataset. Here we benchmark a great number of classical %clasiffiers against our proposed FULL HYBRID enriched with VC-DRC.}.
%\label{Figure 11}
%\end{figure}

\clearpage
%%%%%%%%%%%%%%%%%%%%%%%%%%%%%%%%%%%%%%%%%%%%%%%%%%%%%%%%%%%%%%%%%%%%%%%%%%
%%%%%%%%%%%%%%%%%%%%%%%%%%%%%%%%%%%%%%%%%%%%%%%%%%%%%%%%%%%%%%%%%%%%%%%%%%
\bigskip
\medskip
{\noindent {\em Acknowledgements---}} All the codes used in the manuscript will be provided under the reasonable request. D.A. would like to thank Kishor Bharti for useful discussions on Quantum Machine Learning (QML) and on newly emerging non-VQA algorithms. P.Griffin would like to acknowledge to the Monetary Authority of Singapore (MAS) and to Tradeteq for their support in this work.
\bibliographystyle{apsrev4-1}
\bibliography{IQAE}

\end{document}